\documentclass[%
 reprint,
 amsmath,amssymb,
 aps,
]{revtex4-1}
\usepackage{subfig}
\usepackage{graphicx}
\usepackage{dcolumn}
\usepackage{bm}
\usepackage{epstopdf}
\usepackage[T1]{fontenc}

\begin{document}

\preprint{APS/123-QED}

\title{Superparamagnetism of tryptophan and  walk  memory of proteins}%

\author{Sufi O Raja}

\author{Namrata Jain}
 
 \author{Anjan Kr. Dasgupta}%
 \affiliation{%
 	Department of Biochemistry  \\ University of Calcutta \\ Kolkata 700019 India}
 \altaffiliation[Also at ]{Centre of Excellence in Systems Biology and Biomedical Enineering, University of Calcutta}
 \email{adbioc@caluniv.ac.in}


\begin{abstract}
 Superparamagnetism of tryptophan implying the presence of magnetic domain  is reported. The observation helps us to conceive assembly of proteins as a physical lattice gas with multidimensional Ising character, each lattice points assuming discrete spin states.  When magnetic field is applied the equilibrium is lost and the population density of one spin state increases (unidirectional alignment), resulting in net magnetization. Spatial coherence between the identical spin states further imparts a ferromagnetic memory. This effect is observed using direct nanoscale video imaging. Out of the three proteins ferritin serum albumin and fibrinogen , fibrinogen showed an attenuated response , the protein being essentially  one dimensional. Eventually, Ising lattice is capable of showing ferromagnetic memory only when it has a higher dimensional character. The study highlights possible presence of long range spatial coherence at physiological condition and a plausible microscopic origin of the same. 
. 
\begin{description}

	\item[PACS numbers] {61.48.De,82.60.Cx,82.70.Uv,82.33.Nq}
    \item[Keywords]{\textbf{} Carbon nanomaterials, enthalpy, coupling, surface 
	topology.}
\end{description}
\end{abstract}

\maketitle


\section{\label{sec:level1}Introduction\protect }
Brownian motion of proteins in solution is a subject of recent interest. Correlated random walk of proteins within cells may have functional significance in intra-cellular transport and cell migration \citep{agutter2000random,tabei2013intracellular,wu2014three}. Recently, experimentally it has been shown that the Brownian motions of proteins within live cells shows bias, which is not observed when the same protein is in the solution\cite{di2014probing}. In this paper we studied the effect of external spin (magnetic field) on the random walk of proteins as well as on the sub-cellular streaming pattern. There are reports where effect of magnetic field on diffusion has been studied theoretically. Static magnetic field (SMF) can alter diffusion properties due to occurrence of Lorentzian force and Maxwell stress. Theoretically it has been sown that vortex like motion or constricted phase space (reduced extent of random motion in space) is observed upon magnetic field exposure \cite{kinouchi1988effects,sreenivasan2000evolution}. The differential diffusional property is the outcome of competition between inertial force and induced Lorentzian force. This is the classical view on effect of magnetic field on diffusion. But if the system has an inherent tendency towards correlated random walk , the underlying mechanism is worth exploring from an alternative perspective.\\ Short and long range correlation is an inherent property of Ising lattices \cite{foss2013dynamical}. Ising model has been used to explain the protein aggregation \cite{tsai2014kinetic}, protein folding \cite{liang2003thermodynamics,henry2013comparing}, DNA-protein interaction \cite{etchegoin2003model,teif2010statistical} and very recently cancer metastasis \cite{torquato2011toward}. Amino acids in proteins and proteins in  solution can both be considered as nodes of a 2D Ising lattice at different hierarchical level \cite{garcia2011thermodynamics,hermans1969unfolding,hoang2000spin}. Hence, several emerging properties can be explained based on short or long range correlations within and between the lattices. Recently, extended 1D Ising model has been used to probe the long range correlations within DNA. \\
The present work is  inspired by a question what if there are real spin lattice like behavior of proteins in solution. Our earlier finding \cite{raja2012magnetic} that tryptophan a building block of proteins show memory of static magnetic field. Many of recent interest in quantum mechanical properties of biological molecules also stems from tryptophan distribution \cite{craddock2014feasibility}. 

 Our experimental design is based on probing for a  response to  initial external perturbation by a spin perturbing system like static magnetic field (SMF). As reports of effect of SMF  on biomolecules and cells upon long term (from hours to days) exposure to SMF  are numerous \cite{glade2005brief,freyssinet1983fibrinogen,torbet1984magnetic,chionna2009cell,miyakoshi2005effects,miyakoshi2006review,paul2005strong,teodori2002static,rosen2010studies,higashi1993orientation,hong1995magnetic,tenuzzo2009effect}, the evidence of instant response of towards SMF or the exact mechanism by which live cells translates the magnetic field and competes thermal energy and retains or memorizes such perturbation are rarely discussed.   Our experimental approach seems to provide a basis for existence of a real Ising lattice like behavior of proteins internally in solution and when they are present in cells. \\

\section{\label{sec:level1}Results\protect }

\subsection{\label{sec:level2} Emergence of Magnetism in tryptophan }

 We first show  how the zero-field and with field magnetization alters when we subject tryptophan (trp)a $\pi$-ring  containing amino acid to a pre-exposure to static magnetic field while drying the liquid film of the same. While alteration of its fluorescence emission in response to SMF  was previously reported by our group \cite{raja2012magnetic}, the present work unfolds certain intricate aspects of the magnetic properties of this important (and universal) building block of proteins. Experiments with SQUID ( superconducting quantum interference device) based VSM (Vibrating Sample Magnetometer (VSM) reveals a super paramagnetic behavior of trp. Figure 1 shows the zero-field (solid line) and with field (broken line) magnetization profile of trp at different temperatures.The super paramagnetic nature of the amino acid  is confirmed by presence of a Neel  temperature $T_N$ (indicated by arrow in Figure 1a) \cite{neel1948proprietes} at which $\frac{\delta M}{\delta T}$ changes sign from positive to negative value. The red and blue profiles compare samples  pre-exposed and unexposed to SMF. Incidentally, the existence of $T_N$  for tryptophan ($\approx 100^0K$) asserts its super paramagnetic behavior, this being reported for the first time.  It may be further noted that blocking temperature for field exposed and unexposed samples are identical i.e., $T_{N(SMF exposed)}=T_{N(SMF unexposed)}$.  Now, $T_{N}$ is a product of magnetic anisotropy and domain volume. As magnetization of the exposed sample (zero field, red line) shows a higher value at $T_N$, a higher magnetic anisotropy is expected in the said case, and this has to be  compensated by  reduction of magnetic volume as result of the SMF exposure.\\


\subsection{\label{sec:level2}Magnetic field induced differential walk pattern of proteins in solution }

We performed scattering based imaging using a thin film chamber of three different protein solution before and after magnetic field exposure (see Methods section for experimental details). Figure 2 shows the differential 3D intensity histogram of the population before and after one and five minutes of magnetic field exposure for BSA (Figure 2a, 2b and 2c), ferritin (Figure 2d, 2e and 2f) and fibrinogen (Figure 2g, 2h and 2i). 3D figures represent intensity histogram of individual particle, which has been imaged. A field dependent change in intensity histogram is clear from Figure 2. But the effect is less pronounced for fibrinogen (compare Figure 2a with 2b, 2d with 2e and 2g with 2h). But after five minutes of field exposure change in distribution pattern can be observed (see Figure 2i). \\
To quantify the change observed in Fig. 2,we performed image analysis. We constructed an image plane, which is divided into 10,000 grids (100X100), where X-Y trajectories of the proteins was plotted. See Appendix section for details of the m-code used to construct the image. 
In Fig. 3 the panel A,B,C respectively represents the  X-Y trajectories before exposure  after 1 minute exposure  and after 5 minute exposure to 0.2 Tesla static magnetic field.  The panel D represents the RGB representation of the pixels in A B and C. The presence og white pixels indicate that those segments of the XY  space where the protein motion is independent of field exposure. The figure rows a,b and c of Fig.3 respectively represents studies on BSA, ferritin and fibrinogen. The white pixels are minimal in ferritin and maximal in fibrinogen indicating maximal and minimal magnetic memory in the two respectively. \\ 

The Fig. 3 further implies  that before magnetic field exposure the phase space is isotropic (space filling). A little bit anisotropy in space is observed after 1 minute field exposure and after 5 minutes of exposure the effect become more prominent for BSA and ferritin.  For Fibrinogen the effect is not so prominent, but a rotation in the direction of the trajectories can be clearly observed.  For quantification of the extent anisotropic phase space we calculated an index and assigned as Space Inhomogeneity Index (SII). SII is the ratio of space filled by the proteins and space filled by random distribution of same number of events. Table 1 shows the SII values of proteins before and after magnetic field exposure. The values within the fast bracket indicate the respective drift velocity during video acquisition. \\

\subsection{\label{sec:level2}Magnetic field induced altered sub-cellular streaming }
Figure 4 shows a schematic diagram of the arrangement for confocal microscopic use, in which the instant real time and post  SMF exposure has been studied. As the proteins in solutions showed the remarkable memory effect similar memory effect could be observed in case of intracellular protein movement. In this case both the instant response and the memory effect was studied. We captured six time lapse images of GFP transfected live HeLa cells in absence (Image 1 at t=0 minute and image 2 at t=1 minute), presence (Image 3 at t=2 minute and image 4 at t=3 minute) and after withdrawal (Image 5 at t=4 minute and image 6 at t=5 minute) of the field. The schematics of the magnetic field assisted confocal microscopy and a representative gray scale image of GFP transfected live HeLa cell is given in Appendix section (see Figure S1 and S2). From the six time lapse images we constructed gray scale difference images. Control difference image (see Figure 5a) is constructed by subtraction of image 1 (t=0) from image 2 (t=1). The magnetic effect is observed by construction of difference image (see Figure 5b) through subtraction of image 1 (t=0) from image 4 (t=3). Magnetic memory effect is observed by construction of difference image (see Figure 5c) through subtraction of image 1 (t=0) from image 6 (t=5). Comparison of Figure 1a, 1b and 1c clearly shows that magnetic field exposure results in clustered patchy pixel distribution in the image palne. The effect can also be observed even after the withdrawal of the field.

\section{\label{sec:level3}Discussion\protect }
It was natural to question whether the sttaic field had any effect on the protein structure. Any structural change would lead to changes in the protein walk pattern. Our studies using Circular Dichroism (CD) or ANS fluorescence showed no structural change in the time scale of the study. This indicated that either such changes are absent or the changes if any had an ultra short duration. The reason for the re-orientation of the walk seemed to be the spin-correlated Brownian motion rather than structural change. The long term memory of the walk pattern to such field also implied presence of such correlation rather that structural change in proteins. The spin state of the protein  building block thus affected the overall  protein motion.\\
Next, the direct evidence of alteration in diffusion property (random walk pattern) of proteins by SMF comes from image analysis of the acquired video (see Figure 3). The SMF induced anisotropic phase space can be compared from Table 1. SII value close to 1 means the space is filled homogenously and lower value of SII indicates inhomogeneity in space. That means, SMF induces inhomogeneity in phase space of particle motion (anisotropic movement). From Table 1 it is clear that the magnetic field effect is prominent where the drift velocity is much higher. The initial drift velocity (compare the values within first bracket of Table 1) in case of BSA and Ferritin is very high than Fibrinogen. The effect of SMF is reduction of drift velocity, which can be observed in naked eye from the Video S1, S2 and S3. The effect is evidently amplified if the initial drift velocity is  high. As we imaged the same sample before and after field exposure there is no change in drift velocity due to change in injection flow.
Presence of drift velocity (even under no field condition) indicates presence of an initial  bias in random walk. The particles are moving towards a direction (biased) with their intra-particle random motions. Application of field imparts a  higher Lorentzian force that initially perturbs the spin to a greater extent.Such exposure breaks the bias and results in  an anisotropic streaming of particles in phase space. But, explanation of the memory like effect that is persistence of space anisotropy after field exposure needs to consider SMF induced spin coherence in space. The spatial coherence is a inherent property of evolution of Ising lattice, where memory is a classical output. The two events that seems to control the spin perturbation induced distribution can be summarized as (i) protein shape (ii) drift velocity. Eventually, the physical Ising lattice model explains both the effects.  The shape effect is intrinsically linked with dimensionality of the physical Ising lattice . A chain like protein structure (fibrinogen) may be physically equated to a one dimensional Ising lattice. It is known in literature that one dimensional Ising lattice is devoid of any magnetic memory and can explain only paramagnetic behavior (lack of memory).

In cellular level we also found this type of memory effects. In relaxation image (image without field exposure subtracted from image after withdrawal of the field,i.e, image 6-image 1) we found the retention of the anisotropic subcellular streaming pattern (see Figure 3c). That is exposure of magnetic field is memorized by cells in terms of differential streaming pattern. It is well established that the motion of subcellular proteins is biased. Not only that, but motor like proteins shows Brownian ratchet \cite{peskin1993cellular}like movement, required for cargo transport, where energy for unidirectional movement (bias) comes from ATP hydrolysis \cite{wang2002ratchets,tomkiewicz2007pushing}. Hence, we can say that SMF destroys the inherent biased (correlated) random walk of sub-cellular proteins. The translation of physical force (SMF) is altered sub-cellular streaming, which may be the cause of differential effects on cellular level after long term magnetic incubation. As normal and cancer cell can be considered as different Ising lattices \cite{torquato2011toward}, the onset of magnetic memory in live cell can also be considered as an alternative spin state of the same lattice.The differential effect on static magnetic field on normal and cancer cells  reported by our group \cite{shaw2014modulation}can thus be explained.\\

Ising model has been exploited in explaining various binding processes from last four decades one historically important citation being \cite{changeux1967cooperativity}.  The minimal  Ising lattice, in which the Hamiltonian is given by,
\begin{equation}
H= \sum\nolimits_{<ij>} J_{ij}  \sigma_i \sigma_j -2\beta B\sum\nolimits_{i} \sigma_i 
\end{equation}
In the euqtion (4)  $\sigma$ can have two values $\pm 1$ and the  first term in the LHS represents the coupling between spins that can exist in absence of a field. The second term involving $\beta $ is the contribution due to Bohr magneton that exists only in presence of any external magnetic field B. The  superparamagnetic behavior of tryptophan asserts that the Ising lattice physically exists for this essential protein building block. The presence of a magnetic domain however does not automatically imply emergence of a magnetic memory as one domensional Ising lattice would not exhibit any ferromagnetic memory that would persist in absence of a field. One needs to assume a 2D Ising behaviour whose equations in absence of an external magnetic field are given by ,
\begin{equation}
H= \sum\nolimits_{<ij>} K_{ij}  \sigma_i \sigma_j +\sum\nolimits_{<ij>} L_{ij}  \sigma_i \sigma_j 
\end{equation}
where K and L are vertical and horizontal interaction, one representing the side to side (horizontal) and the other representing a vertical interaction that may originate from say $\pi-\pi$ stacking. Such 2D model with periodic boundary condition may be approximated to a globular protein , where for proteins like fibrinogen the behavior would be more like a one dimensional lattice \\

Once we have the insight in the  magnetic properties the proteins which can be considered as a decorated lattice containing superparamegnetic domains. We can then look upon the protein walk as a Brownian motion of superparamagnetic particles \cite{yue2012motion,chen2013magnetic}. While the authors in these work did not consider the memory of an initial magnetic perturbation, their analysis highlights importance of Brownian motion of superparamagnetic particles in context of drug targeting and  designed magnetic arrays(traps). Our findings highlight that even  proteins can be subjected to magnetically controlled behavior,  the memory component of the magnetic contribution depending on the protein shape and dimensionality. It may be noted that simple diamagnetic properties would not suffice to explain the memory aspect. \\

It also follows from Table 1  that the random walk seems to have least spin perturbation memory for Fibrinogen , whereas for albumin and ferritin there seems to be a significant random walk shift. The space filling index  provides a good estimate of the   memory. Interestingly, the simple numerical expression  can be applied to determine the relative magnetic moments of ferrofluids and other magnetic nanomaterials.  \\
The observation may have important implications in another context. The presence  super paramagnetism  in the individual amino acid scale and emergence of a ferromagnetic like memory manifested in the random walk in solutions (and also in protein motion in cells) implies presence of a temporally stable long term spatial coherence in physiological conditions. In quantum biology some approaches have been sensing weak field \cite{lambert2013quantum,bhattacharya2014static} and events triggered by extreme small scale temporal rhythms often termed as quantum beats \cite{panitchayangkoon2010long}.Presence of large spatial coherence that is reflected in random walk for a wide variety of proteins, has rarely been discussed. The observation may therefore be contextual with respect to deciphering many cellular communication processes , where very raraely the importance of spatial coherence is discussed. Very recently importance of protein dynamics has been emphasized in modification process like ubiquitination and protein-protein interaction. The presence of local spatial coherence may modulate or may open up control of such interactions.

\section{Conclusion}
The proteins movements \textit{in vitro }(i.e in solution ) and \textit{in vivo} (i.e within the cytoplasm ) show a field memory, this being  reported for the first time. Mechanistically the behavior is explained by the finding that the building block of proteins ,e.g. its aromatic amino acid tryptophan may have super paramagnetic behavior. This inspired a direct exploitation of the  Ising model , the previous Ising model based studies on biomolecular interactions \cite{changeux1967cooperativity} being primarily analogy based.


%
%

\begin{acknowledgments}
The authors wish to thank Dr. S.Chattpadhyay for the SQUID /VSM experiment at CRNN. 
\end{acknowledgments}

\appendix

\section{Appendixes}

\subsection{Materials and Methods}

	L-tryptophan, Ferritin and Fibrinogen were purchased from Sigma. The Bovine Serum Albumin (BSA) was purchased from SRL. Millipore water was used to prepare stock solution and dilution. Protein solutions were prepared using 100 mM phosphate buffer of pH 7.22. 20 mg L-tryptophan was dissolved in 10 ml of MilliQ water.\\
	The permanent Neodymium magnets were purchased from Rare Earth Magnetics, India. The field strength was measured using a locally made Gaussmeter (Neoequipments, India).\\
	\section*{Magnetic measurement}
	The prepared 10 ml trytophan solution was divided into two parts and dried using a lyophillizer, one part in absence of magnetic field and other one in presence of 0.2 Tesla magnetic field. Then the magnetic measurement was performed in Vibrational Sample Magnetometer (Quantum Design, MPMS 7). Both samples were heated to 300K in presence of 100 Oesterd field and then cooling up to 4K was performed in presence (field cooling) and in absence (zero field cooling) of magnetic field. Magnetization vs Field (M-H) study was performed using oscillatory field from -10,000 to +10,000 Oe externally applied magnetic field.
	
	\section*{Nanoparticle Tracking Analysis (NTA)}
	NTA measurements were performed using NanoSight NS300 (Malvern, United Kingdom) system equipped with a high sensitivity CMOS camera and 532 nm laser. All	measurements were performed at room temperature.The software used for capturing and analyzing the data was the NTA 2.3. 0.1 mg/ml protein solution was imaged before, after 1 minute and after 5 minutes of magnetic field exposure. The video imaging was performed for 1 minute for each condition at 24 frames per second rate.\\
	\section*{Transfection of HeLa cells}
	HeLa cell lines (passage no: 15) was procurred from NCCS, India. Cells were cultured in DMEM, high glucose (Gibco 12800-058) with 10 percent heat inactivated FBS and the media is supplemented with 1X antibiotic (PenStrep Gibco). Cells were maintained at 5 percent CO2 (Carbon di Oxide) environment. For subsequent passages, cells were dislodged using 0.25 percent Trypsin EDTA (1X) (GIBCO 25200). HeLa cells were transfected by pEGFP-N3 plasmids using calcium phosphate transfection protocol as described previously \cite{jordan1996transfecting}.
	\section*{Magnetic field assisted confocal microscopy} 
	Time lapse confocal imaging was performed in Olympus (FV1000) microscope. Six time lapse images were acquired at 1 minute time interval. First and second image were captured in absence of magnetic field. Third and fourth image were captured in presence of magnetic field. The fifth and sixth image were captured after withdrawal of the magnetic field. The schematic illustration of the magnetic field assisted microscopy is provided in the Supplementary section. \\


M-code

\subsection{Algorithm}
\begin{verbatim*}
function z=raja_grid(x,y,m,n);
% m & n are sizes of the matrix
ik=find(x<0);
x(ik)=[];
y(ik)=[];
ik=find(y<0);
x(ik)=[];
y(ik)=[];
% x=x/max(x);
% y=y/max(y);
z=zeros(m,n);
xl=linspace(min(x),max(x),m);
yl=linspace(min(y),max(y),n);


for i=1:m-1;
for j=1:n-1
kk=find (x>=xl(i) & x<=xl(i+1) & y>=yl(j)  & y<=yl(j+1) );
%if isempty(kk)~=1 

%z(i,j)=z(i,j)+1;
%z(i,j)=255*(sum(I(kk))./max(I(kk)));
%end
end
end
z=uint8(z);
w(:,:,1)=z;w(:,:,2)=z;w(:,:,3)=z;

%w is the gray scale output image.
\end{verbatim*}

\pagebreak

\begin{table*}
	\caption{Space Inhomogeneity Index $(SII)$ for three different proteins and reduction of drift velocity of proteins upon magnetic field exposure, percentage ratio of space filled by the particle motion (observation)/ space filled by random distribution of same number of particles (simulation). the values within the first bracket is the respective drift velocity in $\mu meter sec^{-1}$.}
	\begin{center}
		\begin{tabular}{|l|l|l|l|}
			\hline
			Proteins& 
			SII (unexposed)& 
			SII (exposed 1)& 
			SII (exposed 2) \\
			\hline
			BSA& 
			0.862 (11.718)& 
			0.686 (3.101)& 
			0.585 (0.612) \\
			\hline
			Ferritin& 
			0.911 (20.045)& 
			0.528 (2.894)& 
			0.337 (0. 284) \\
			\hline
			Fibrinogen& 
			0.855 (9.548)& 
			0.852 (9.62)& 
			0.789 (4.812) \\
			\hline
		\end{tabular}
		
		\label{tab1}
	\end{center}
\end{table*}

\begin{figure*}[htbp]
	\includegraphics[width =7in ,height=4in]{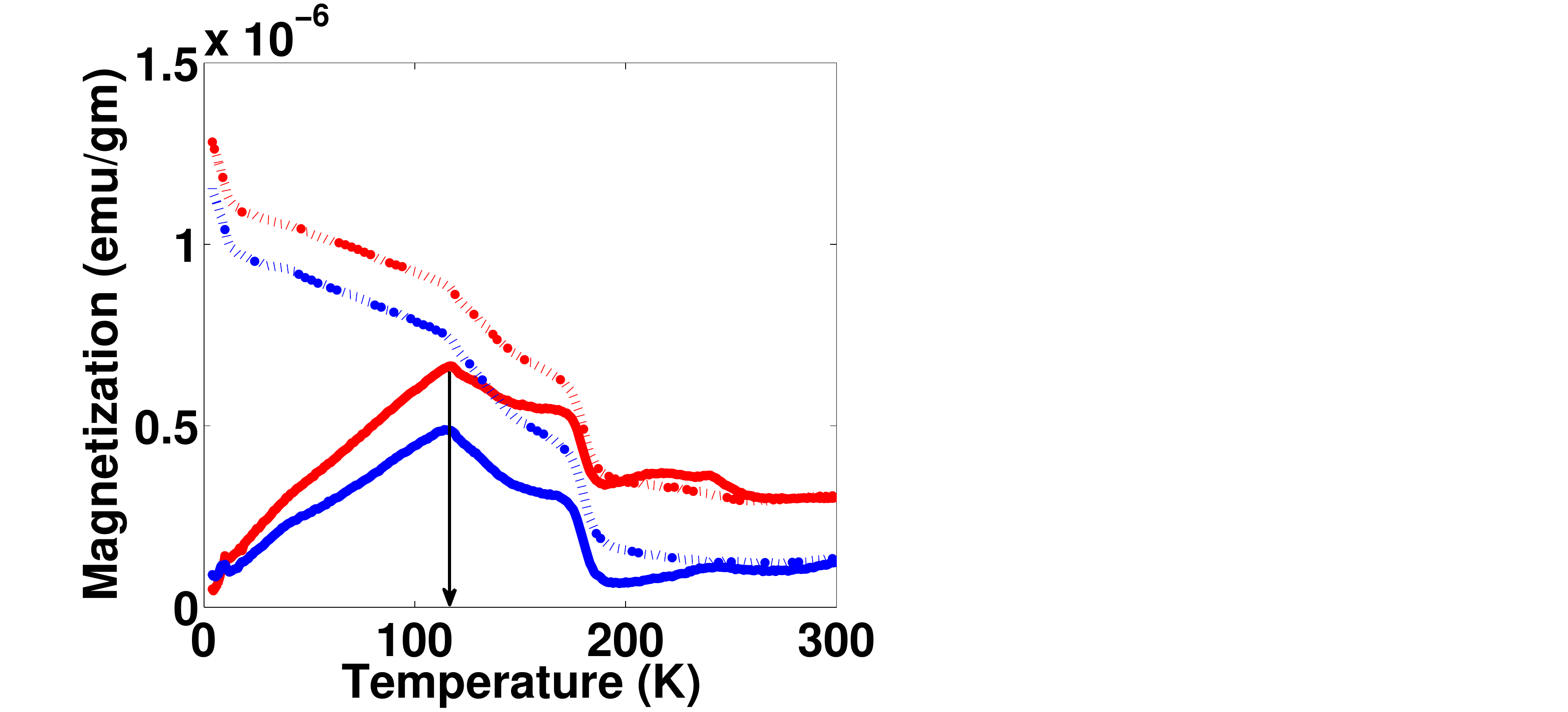}
	\label{fg:zfc}
	\caption{Superparamagnetic property of tryptophan -  Magnetization vs temperature (M-T) measurement of SMF preincubated (red lines) and control (blue lines) tryptophan. Solid and broken lines indicates Zero Field Cooling (ZFC) and Field Cooling (FC) profiles}
	
\end{figure*}

\begin{figure*}[htbp]
	\includegraphics[width =7in ,height=4in]{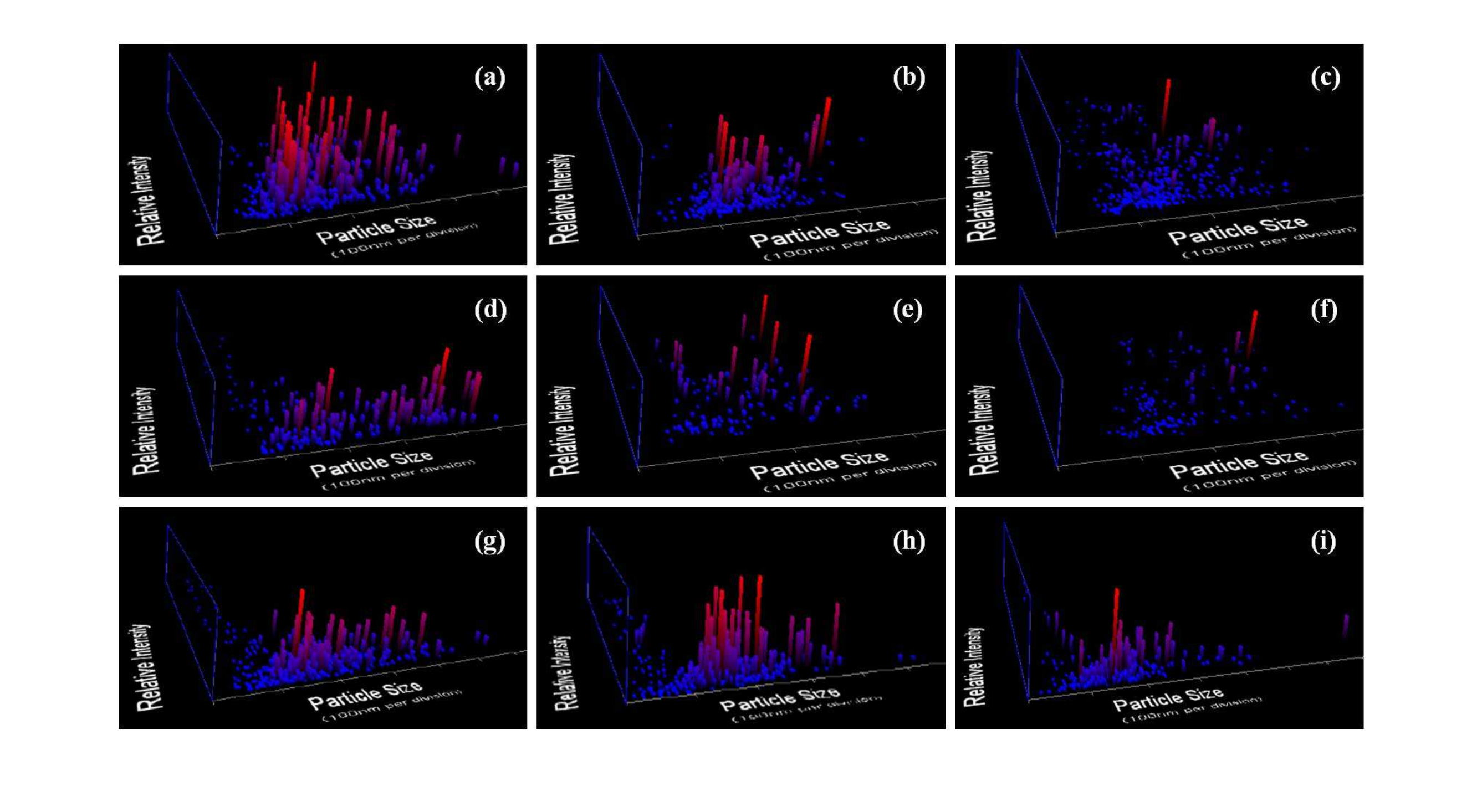}
	\label{fg:histogram}
	\caption{The upper middle and lower rows respectively represents population distribution for  BSA Ferritin  and Fibrinogen respectively. The left middle and right columns respectively represents distribution  before  field exposure, after  1 minute   and after  and 5 minutes exposures to 0.2 Tesla magnetic field.}
	
\end{figure*}


\begin{figure*}[ht]
	
	\centering
	\subfloat[]{\includegraphics[height = 0.4\textwidth,width=0.8\textwidth]{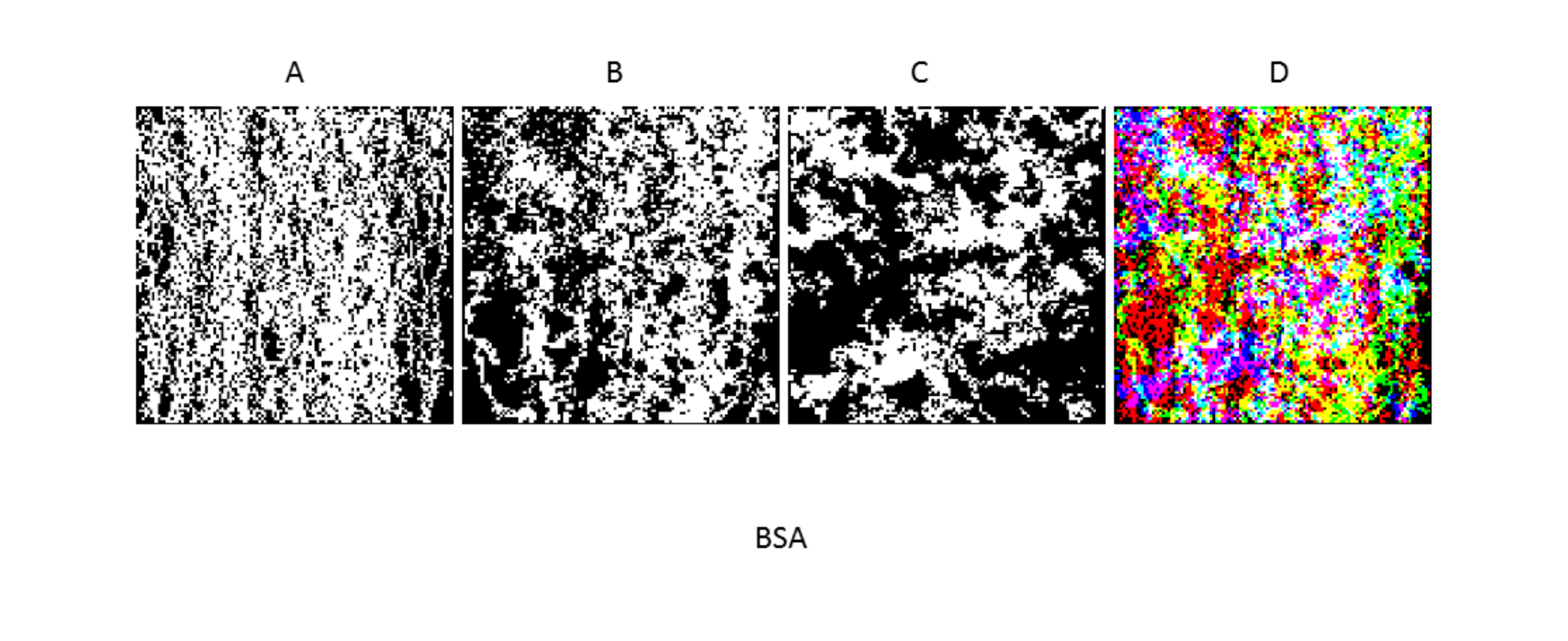}\label{fg:bsa}}\\
	\subfloat[]{\includegraphics[height = 0.4\textwidth,width=0.8\textwidth]{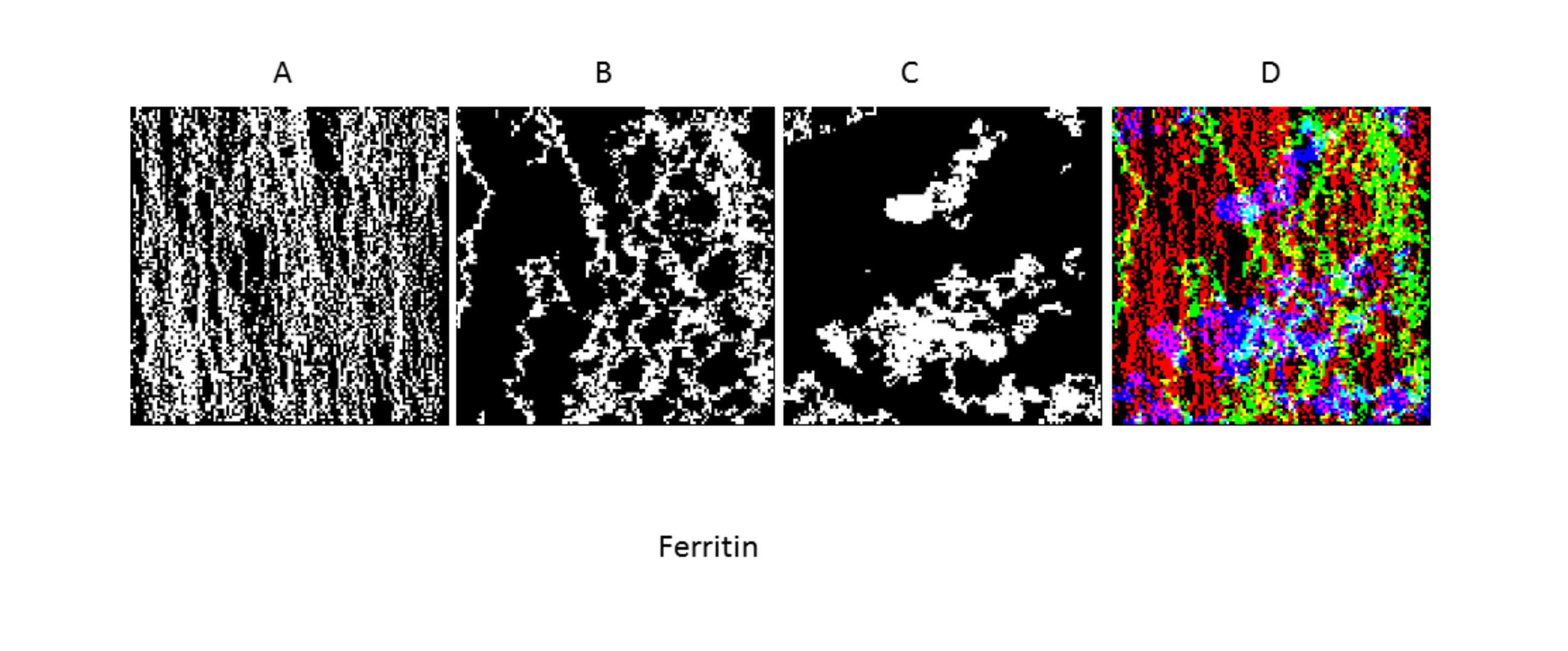}\label{fg:ferritin}}\\
	\subfloat[]{\includegraphics[height = 0.4\textwidth,width=0.8\textwidth]{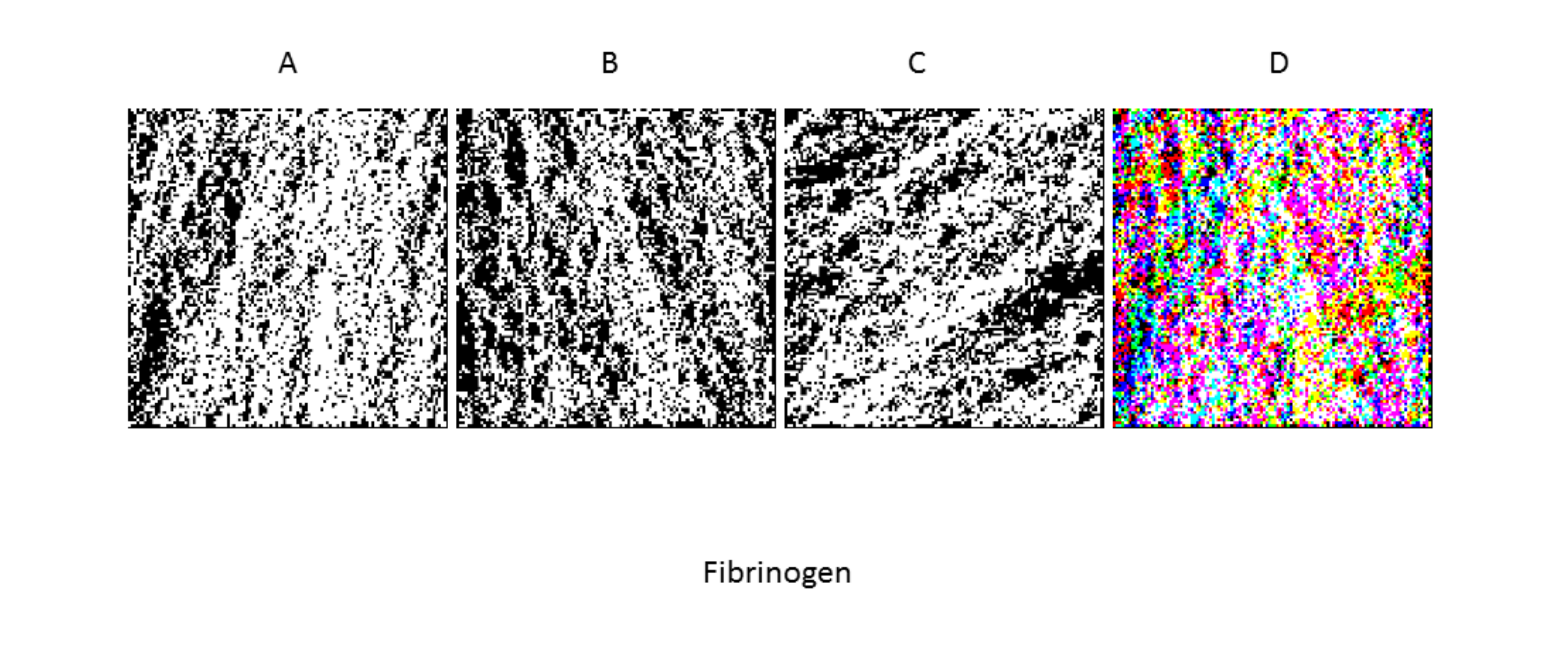}\label{fg:fibrinogen}}\\
	\caption{Protein Walk - Direct visualization A, B , C are states exposed to no exposure and 1 min exposure and 5 min exposure to 0.2T field. D represent fusion profile generated assuminmg A,B,C to be located at the R,G and B planes. {(a) BSA (b) ferritin (c)fibrinogen}- white pixels in the D panel of fibrinogen indicates attenuated response of the protein to SMF}
\label{fg:allvideo}
\end{figure*}
\begin{figure*}[htbp]
	\includegraphics[width =6in ,height=2.5in]{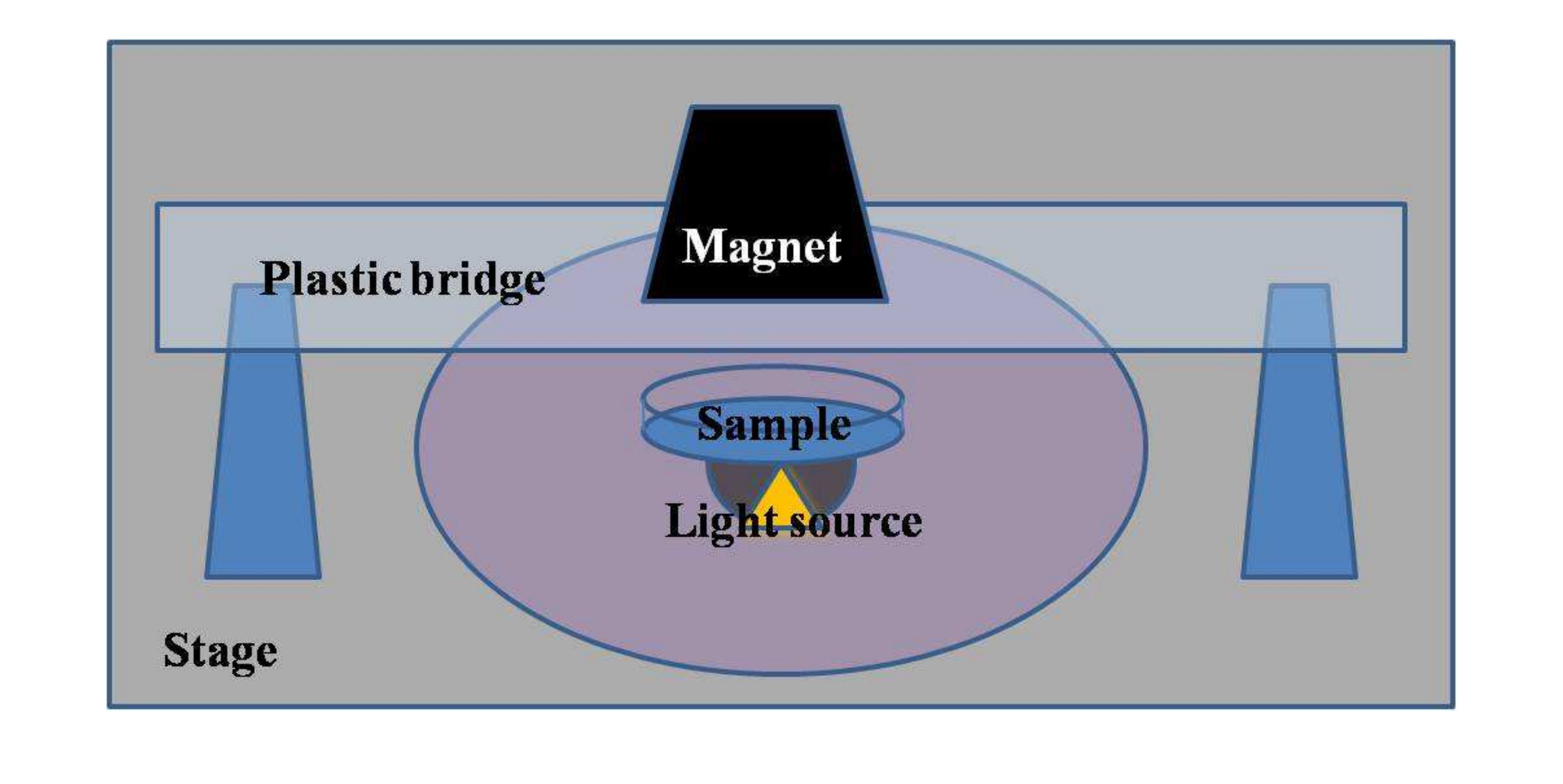}
		\label{fg:schematic}
	\caption{Schematic illustration of the design of magnetic field assisted confocal microscopy}

\end{figure*}

\begin{figure*}[htbp]
		\includegraphics[width =6in ,height=2.5in]{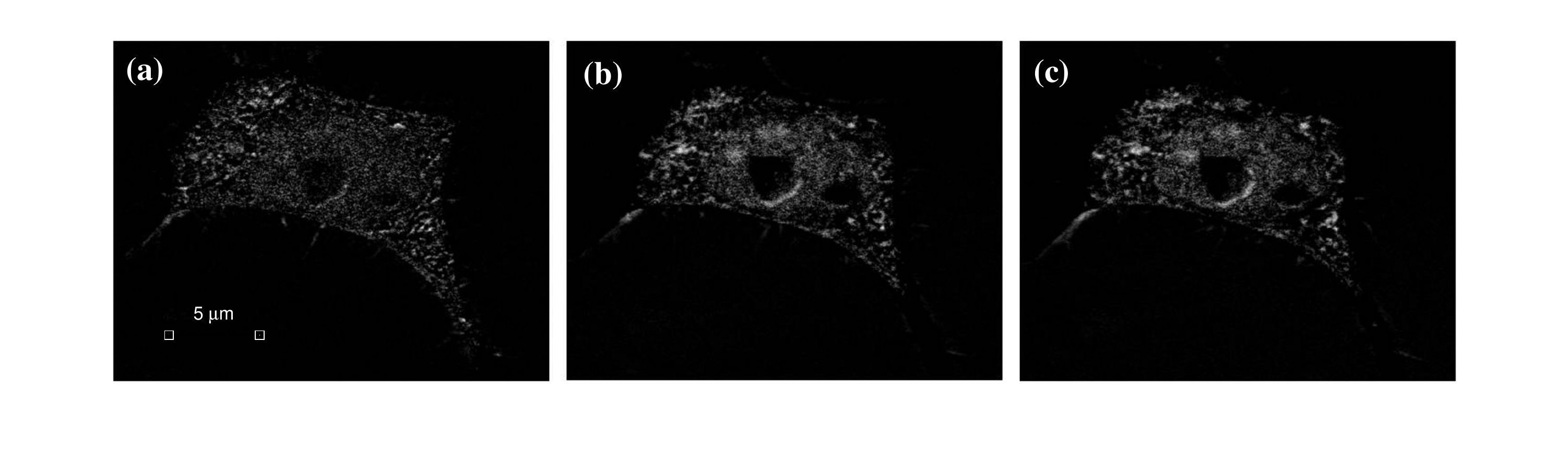}
	\caption{Magnetic field induced altered walk patterns proteins at the subcellular level. Six time lapse images of GFP transfected live HeLa cell were captured at t=0, t=1, t=2, t=3, t=4 and t=5 minute. At t=0 and 1 magnetic field was absent, at t=2 and 3 magnetic field was present and at t=4 and 5 the field was withdrawn.(a), (b) and (c) are the gray scale difference image. (a) is the control difference image [image at t(1)-image at t(0)], (b) is the magnetic difference image [image at t(3)-image at t(0)] and (c) is the magnetic memory image [image at t(5)-image at t(0)].The original gray scale confocal image is provided in Supplementary section}
	\label{fg:video}
\end{figure*}

%


\begin{thebibliography}{42}%
	\makeatletter
	\providecommand \@ifxundefined [1]{%
		\@ifx{#1\undefined}
	}%
	\providecommand \@ifnum [1]{%
		\ifnum #1\expandafter \@firstoftwo
		\else \expandafter \@secondoftwo
		\fi
	}%
	\providecommand \@ifx [1]{%
		\ifx #1\expandafter \@firstoftwo
		\else \expandafter \@secondoftwo
		\fi
	}%
	\providecommand \natexlab [1]{#1}%
	\providecommand \enquote  [1]{``#1''}%
	\providecommand \bibnamefont  [1]{#1}%
	\providecommand \bibfnamefont [1]{#1}%
	\providecommand \citenamefont [1]{#1}%
	\providecommand \href@noop [0]{\@secondoftwo}%
	\providecommand \href [0]{\begingroup \@sanitize@url \@href}%
	\providecommand \@href[1]{\@@startlink{#1}\@@href}%
	\providecommand \@@href[1]{\endgroup#1\@@endlink}%
	\providecommand \@sanitize@url [0]{\catcode `\\12\catcode `\$12\catcode
		`\&12\catcode `\#12\catcode `\^12\catcode `\_12\catcode `\%12\relax}%
	\providecommand \@@startlink[1]{}%
	\providecommand \@@endlink[0]{}%
	\providecommand \url  [0]{\begingroup\@sanitize@url \@url }%
	\providecommand \@url [1]{\endgroup\@href {#1}{\urlprefix }}%
	\providecommand \urlprefix  [0]{URL }%
	\providecommand \Eprint [0]{\href }%
	\providecommand \doibase [0]{http://dx.doi.org/}%
	\providecommand \selectlanguage [0]{\@gobble}%
	\providecommand \bibinfo  [0]{\@secondoftwo}%
	\providecommand \bibfield  [0]{\@secondoftwo}%
	\providecommand \translation [1]{[#1]}%
	\providecommand \BibitemOpen [0]{}%
	\providecommand \bibitemStop [0]{}%
	\providecommand \bibitemNoStop [0]{.\EOS\space}%
	\providecommand \EOS [0]{\spacefactor3000\relax}%
	\providecommand \BibitemShut  [1]{\csname bibitem#1\endcsname}%
	\let\auto@bib@innerbib\@empty
	\bibitem [{\citenamefont {Agutter}\ and\ \citenamefont
		{Wheatley}(2000)}]{agutter2000random}%
	\BibitemOpen
	\bibfield  {author} {\bibinfo {author} {\bibfnamefont {P.~S.}\ \bibnamefont
			{Agutter}}\ and\ \bibinfo {author} {\bibfnamefont {D.~N.}\ \bibnamefont
			{Wheatley}},\ }\href@noop {} {\bibfield  {journal} {\bibinfo  {journal}
			{BioEssays}\ }\textbf {\bibinfo {volume} {22}},\ \bibinfo {pages} {1018}
		(\bibinfo {year} {2000})}\BibitemShut {NoStop}%
	\bibitem [{\citenamefont {Tabei}\ \emph {et~al.}(2013)\citenamefont {Tabei},
		\citenamefont {Burov}, \citenamefont {Kim}, \citenamefont {Kuznetsov},
		\citenamefont {Huynh}, \citenamefont {Jureller}, \citenamefont {Philipson},
		\citenamefont {Dinner},\ and\ \citenamefont
		{Scherer}}]{tabei2013intracellular}%
	\BibitemOpen
	\bibfield  {author} {\bibinfo {author} {\bibfnamefont {S.~A.}\ \bibnamefont
			{Tabei}}, \bibinfo {author} {\bibfnamefont {S.}~\bibnamefont {Burov}},
		\bibinfo {author} {\bibfnamefont {H.~Y.}\ \bibnamefont {Kim}}, \bibinfo
		{author} {\bibfnamefont {A.}~\bibnamefont {Kuznetsov}}, \bibinfo {author}
		{\bibfnamefont {T.}~\bibnamefont {Huynh}}, \bibinfo {author} {\bibfnamefont
			{J.}~\bibnamefont {Jureller}}, \bibinfo {author} {\bibfnamefont {L.~H.}\
			\bibnamefont {Philipson}}, \bibinfo {author} {\bibfnamefont {A.~R.}\
			\bibnamefont {Dinner}}, \ and\ \bibinfo {author} {\bibfnamefont {N.~F.}\
			\bibnamefont {Scherer}},\ }\href@noop {} {\bibfield  {journal} {\bibinfo
			{journal} {Proceedings of the National Academy of Sciences}\ }\textbf
		{\bibinfo {volume} {110}},\ \bibinfo {pages} {4911} (\bibinfo {year}
		{2013})}\BibitemShut {NoStop}%
	\bibitem [{\citenamefont {Wu}\ \emph {et~al.}(2014)\citenamefont {Wu},
		\citenamefont {Giri}, \citenamefont {Sun},\ and\ \citenamefont
		{Wirtz}}]{wu2014three}%
	\BibitemOpen
	\bibfield  {author} {\bibinfo {author} {\bibfnamefont {P.-H.}\ \bibnamefont
			{Wu}}, \bibinfo {author} {\bibfnamefont {A.}~\bibnamefont {Giri}}, \bibinfo
		{author} {\bibfnamefont {S.~X.}\ \bibnamefont {Sun}}, \ and\ \bibinfo
		{author} {\bibfnamefont {D.}~\bibnamefont {Wirtz}},\ }\href@noop {}
	{\bibfield  {journal} {\bibinfo  {journal} {Proceedings of the National
				Academy of Sciences}\ }\textbf {\bibinfo {volume} {111}},\ \bibinfo {pages}
		{3949} (\bibinfo {year} {2014})}\BibitemShut {NoStop}%
	\bibitem [{\citenamefont {Di~Rienzo}\ \emph {et~al.}(2014)\citenamefont
		{Di~Rienzo}, \citenamefont {Piazza}, \citenamefont {Gratton}, \citenamefont
		{Beltram},\ and\ \citenamefont {Cardarelli}}]{di2014probing}%
	\BibitemOpen
	\bibfield  {author} {\bibinfo {author} {\bibfnamefont {C.}~\bibnamefont
			{Di~Rienzo}}, \bibinfo {author} {\bibfnamefont {V.}~\bibnamefont {Piazza}},
		\bibinfo {author} {\bibfnamefont {E.}~\bibnamefont {Gratton}}, \bibinfo
		{author} {\bibfnamefont {F.}~\bibnamefont {Beltram}}, \ and\ \bibinfo
		{author} {\bibfnamefont {F.}~\bibnamefont {Cardarelli}},\ }\href@noop {}
	{\bibfield  {journal} {\bibinfo  {journal} {Nature communications}\ }\textbf
		{\bibinfo {volume} {5}} (\bibinfo {year} {2014})}\BibitemShut {NoStop}%
	\bibitem [{\citenamefont {Kinouchi}\ \emph {et~al.}(1988)\citenamefont
		{Kinouchi}, \citenamefont {Tanimoto}, \citenamefont {Ushita}, \citenamefont
		{Sato}, \citenamefont {Yamaguchi},\ and\ \citenamefont
		{Miyamoto}}]{kinouchi1988effects}%
	\BibitemOpen
	\bibfield  {author} {\bibinfo {author} {\bibfnamefont {Y.}~\bibnamefont
			{Kinouchi}}, \bibinfo {author} {\bibfnamefont {S.}~\bibnamefont {Tanimoto}},
		\bibinfo {author} {\bibfnamefont {T.}~\bibnamefont {Ushita}}, \bibinfo
		{author} {\bibfnamefont {K.}~\bibnamefont {Sato}}, \bibinfo {author}
		{\bibfnamefont {H.}~\bibnamefont {Yamaguchi}}, \ and\ \bibinfo {author}
		{\bibfnamefont {H.}~\bibnamefont {Miyamoto}},\ }\href@noop {} {\bibfield
		{journal} {\bibinfo  {journal} {Bioelectromagnetics}\ }\textbf {\bibinfo
			{volume} {9}},\ \bibinfo {pages} {159} (\bibinfo {year} {1988})}\BibitemShut
	{NoStop}%
	\bibitem [{\citenamefont {Sreenivasan}\ and\ \citenamefont
		{Alboussi{\`e}re}(2000)}]{sreenivasan2000evolution}%
	\BibitemOpen
	\bibfield  {author} {\bibinfo {author} {\bibfnamefont {B.}~\bibnamefont
			{Sreenivasan}}\ and\ \bibinfo {author} {\bibfnamefont {T.}~\bibnamefont
			{Alboussi{\`e}re}},\ }\href@noop {} {\bibfield  {journal} {\bibinfo
			{journal} {European Journal of Mechanics-B/Fluids}\ }\textbf {\bibinfo
			{volume} {19}},\ \bibinfo {pages} {403} (\bibinfo {year} {2000})}\BibitemShut
	{NoStop}%
	\bibitem [{\citenamefont {Foss-Feig}\ \emph {et~al.}(2013)\citenamefont
		{Foss-Feig}, \citenamefont {Hazzard}, \citenamefont {Bollinger},
		\citenamefont {Rey},\ and\ \citenamefont {Clark}}]{foss2013dynamical}%
	\BibitemOpen
	\bibfield  {author} {\bibinfo {author} {\bibfnamefont {M.}~\bibnamefont
			{Foss-Feig}}, \bibinfo {author} {\bibfnamefont {K.~R.}\ \bibnamefont
			{Hazzard}}, \bibinfo {author} {\bibfnamefont {J.~J.}\ \bibnamefont
			{Bollinger}}, \bibinfo {author} {\bibfnamefont {A.~M.}\ \bibnamefont {Rey}},
		\ and\ \bibinfo {author} {\bibfnamefont {C.~W.}\ \bibnamefont {Clark}},\
	}\href@noop {} {\bibfield  {journal} {\bibinfo  {journal} {New Journal of
			Physics}\ }\textbf {\bibinfo {volume} {15}},\ \bibinfo {pages} {113008}
	(\bibinfo {year} {2013})}\BibitemShut {NoStop}%
\bibitem [{\citenamefont {Tsai}\ \emph {et~al.}(2014)\citenamefont {Tsai},
	\citenamefont {Yuan},\ and\ \citenamefont {Lin}}]{tsai2014kinetic}%
\BibitemOpen
\bibfield  {author} {\bibinfo {author} {\bibfnamefont {M.-Y.}\ \bibnamefont
		{Tsai}}, \bibinfo {author} {\bibfnamefont {J.-M.}\ \bibnamefont {Yuan}}, \
	and\ \bibinfo {author} {\bibfnamefont {S.-H.}\ \bibnamefont {Lin}},\
}\href@noop {} {\bibfield  {journal} {\bibinfo  {journal} {Biophysical
		Journal}\ }\textbf {\bibinfo {volume} {106}},\ \bibinfo {pages} {59a}
(\bibinfo {year} {2014})}\BibitemShut {NoStop}%
\bibitem [{\citenamefont {Liang}\ \emph {et~al.}(2003)\citenamefont {Liang},
	\citenamefont {Hayashi}, \citenamefont {Shiu}, \citenamefont {Mo},
	\citenamefont {Shao}, \citenamefont {Yan},\ and\ \citenamefont
	{Lin}}]{liang2003thermodynamics}%
\BibitemOpen
\bibfield  {author} {\bibinfo {author} {\bibfnamefont {K.~K.}\ \bibnamefont
		{Liang}}, \bibinfo {author} {\bibfnamefont {M.}~\bibnamefont {Hayashi}},
	\bibinfo {author} {\bibfnamefont {Y.}~\bibnamefont {Shiu}}, \bibinfo {author}
	{\bibfnamefont {Y.}~\bibnamefont {Mo}}, \bibinfo {author} {\bibfnamefont
		{J.}~\bibnamefont {Shao}}, \bibinfo {author} {\bibfnamefont {Y.}~\bibnamefont
		{Yan}}, \ and\ \bibinfo {author} {\bibfnamefont {S.~H.}\ \bibnamefont
		{Lin}},\ }\href@noop {} {\bibfield  {journal} {\bibinfo  {journal} {Physical
			Chemistry Chemical Physics}\ }\textbf {\bibinfo {volume} {5}},\ \bibinfo
	{pages} {5300} (\bibinfo {year} {2003})}\BibitemShut {NoStop}%
\bibitem [{\citenamefont {Henry}\ \emph {et~al.}(2013)\citenamefont {Henry},
	\citenamefont {Best},\ and\ \citenamefont {Eaton}}]{henry2013comparing}%
\BibitemOpen
\bibfield  {author} {\bibinfo {author} {\bibfnamefont {E.~R.}\ \bibnamefont
		{Henry}}, \bibinfo {author} {\bibfnamefont {R.~B.}\ \bibnamefont {Best}}, \
	and\ \bibinfo {author} {\bibfnamefont {W.~A.}\ \bibnamefont {Eaton}},\
}\href@noop {} {\bibfield  {journal} {\bibinfo  {journal} {Proceedings of the
		National Academy of Sciences}\ }\textbf {\bibinfo {volume} {110}},\ \bibinfo
{pages} {17880} (\bibinfo {year} {2013})}\BibitemShut {NoStop}%
\bibitem [{\citenamefont {Etchegoin}\ and\ \citenamefont
	{N{\"o}llmann}(2003)}]{etchegoin2003model}%
\BibitemOpen
\bibfield  {author} {\bibinfo {author} {\bibfnamefont {P.}~\bibnamefont
		{Etchegoin}}\ and\ \bibinfo {author} {\bibfnamefont {M.}~\bibnamefont
		{N{\"o}llmann}},\ }\href@noop {} {\bibfield  {journal} {\bibinfo  {journal}
		{Journal of theoretical biology}\ }\textbf {\bibinfo {volume} {220}},\
	\bibinfo {pages} {233} (\bibinfo {year} {2003})}\BibitemShut {NoStop}%
\bibitem [{\citenamefont {Teif}\ and\ \citenamefont
	{Rippe}(2010)}]{teif2010statistical}%
\BibitemOpen
\bibfield  {author} {\bibinfo {author} {\bibfnamefont {V.~B.}\ \bibnamefont
		{Teif}}\ and\ \bibinfo {author} {\bibfnamefont {K.}~\bibnamefont {Rippe}},\
}\href@noop {} {\bibfield  {journal} {\bibinfo  {journal} {Journal of
		Physics: Condensed Matter}\ }\textbf {\bibinfo {volume} {22}},\ \bibinfo
{pages} {414105} (\bibinfo {year} {2010})}\BibitemShut {NoStop}%
\bibitem [{\citenamefont {Torquato}(2011)}]{torquato2011toward}%
\BibitemOpen
\bibfield  {author} {\bibinfo {author} {\bibfnamefont {S.}~\bibnamefont
		{Torquato}},\ }\href@noop {} {\bibfield  {journal} {\bibinfo  {journal}
		{Physical biology}\ }\textbf {\bibinfo {volume} {8}},\ \bibinfo {pages}
	{015017} (\bibinfo {year} {2011})}\BibitemShut {NoStop}%
\bibitem [{\citenamefont {Garcia}\ \emph {et~al.}(2011)\citenamefont {Garcia},
	\citenamefont {Kondev}, \citenamefont {Orme}, \citenamefont {Theriot},\ and\
	\citenamefont {Phillips}}]{garcia2011thermodynamics}%
\BibitemOpen
\bibfield  {author} {\bibinfo {author} {\bibfnamefont {H.~G.}\ \bibnamefont
		{Garcia}}, \bibinfo {author} {\bibfnamefont {J.}~\bibnamefont {Kondev}},
	\bibinfo {author} {\bibfnamefont {N.}~\bibnamefont {Orme}}, \bibinfo {author}
	{\bibfnamefont {J.~A.}\ \bibnamefont {Theriot}}, \ and\ \bibinfo {author}
	{\bibfnamefont {R.}~\bibnamefont {Phillips}},\ }\href@noop {} {\bibfield
	{journal} {\bibinfo  {journal} {Methods in enzymology}\ }\textbf {\bibinfo
		{volume} {492}},\ \bibinfo {pages} {27} (\bibinfo {year} {2011})}\BibitemShut
{NoStop}%
\bibitem [{\citenamefont {Hermans}\ \emph {et~al.}(1969)\citenamefont
	{Hermans}, \citenamefont {Lohr},\ and\ \citenamefont
	{Ferro}}]{hermans1969unfolding}%
\BibitemOpen
\bibfield  {author} {\bibinfo {author} {\bibfnamefont {J.}~\bibnamefont
		{Hermans}}, \bibinfo {author} {\bibfnamefont {D.}~\bibnamefont {Lohr}}, \
	and\ \bibinfo {author} {\bibfnamefont {D.}~\bibnamefont {Ferro}},\
}\href@noop {} {\  (\bibinfo {year} {1969})}\BibitemShut {NoStop}%
\bibitem [{\citenamefont {Hoang}\ \emph {et~al.}(2000)\citenamefont {Hoang},
	\citenamefont {Sushko}, \citenamefont {Li},\ and\ \citenamefont
	{Cieplak}}]{hoang2000spin}%
\BibitemOpen
\bibfield  {author} {\bibinfo {author} {\bibfnamefont {T.~X.}\ \bibnamefont
		{Hoang}}, \bibinfo {author} {\bibfnamefont {N.}~\bibnamefont {Sushko}},
	\bibinfo {author} {\bibfnamefont {M.~S.}\ \bibnamefont {Li}}, \ and\ \bibinfo
	{author} {\bibfnamefont {M.}~\bibnamefont {Cieplak}},\ }\href@noop {}
{\bibfield  {journal} {\bibinfo  {journal} {Journal of Physics A:
			Mathematical and General}\ }\textbf {\bibinfo {volume} {33}},\ \bibinfo
	{pages} {3977} (\bibinfo {year} {2000})}\BibitemShut {NoStop}%
\bibitem [{\citenamefont {Raja}\ and\ \citenamefont
	{Dasgupta}(2012)}]{raja2012magnetic}%
\BibitemOpen
\bibfield  {author} {\bibinfo {author} {\bibfnamefont {S.~O.}\ \bibnamefont
		{Raja}}\ and\ \bibinfo {author} {\bibfnamefont {A.~K.}\ \bibnamefont
		{Dasgupta}},\ }\href@noop {} {\bibfield  {journal} {\bibinfo  {journal}
		{Chemical Physics Letters}\ }\textbf {\bibinfo {volume} {554}},\ \bibinfo
	{pages} {163} (\bibinfo {year} {2012})}\BibitemShut {NoStop}%
\bibitem [{\citenamefont {Craddock}\ \emph {et~al.}(2014)\citenamefont
	{Craddock}, \citenamefont {Friesen}, \citenamefont {Mane}, \citenamefont
	{Hameroff},\ and\ \citenamefont {Tuszynski}}]{craddock2014feasibility}%
\BibitemOpen
\bibfield  {author} {\bibinfo {author} {\bibfnamefont {T.~J.~A.}\
		\bibnamefont {Craddock}}, \bibinfo {author} {\bibfnamefont {D.}~\bibnamefont
		{Friesen}}, \bibinfo {author} {\bibfnamefont {J.}~\bibnamefont {Mane}},
	\bibinfo {author} {\bibfnamefont {S.}~\bibnamefont {Hameroff}}, \ and\
	\bibinfo {author} {\bibfnamefont {J.~A.}\ \bibnamefont {Tuszynski}},\
}\href@noop {} {\bibfield  {journal} {\bibinfo  {journal} {Journal of The
		Royal Society Interface}\ }\textbf {\bibinfo {volume} {11}},\ \bibinfo
{pages} {20140677} (\bibinfo {year} {2014})}\BibitemShut {NoStop}%
\bibitem [{\citenamefont {Glade}\ and\ \citenamefont
	{Tabony}(2005)}]{glade2005brief}%
\BibitemOpen
\bibfield  {author} {\bibinfo {author} {\bibfnamefont {N.}~\bibnamefont
		{Glade}}\ and\ \bibinfo {author} {\bibfnamefont {J.}~\bibnamefont {Tabony}},\
}\href@noop {} {\bibfield  {journal} {\bibinfo  {journal} {Biophysical
		chemistry}\ }\textbf {\bibinfo {volume} {115}},\ \bibinfo {pages} {29}
(\bibinfo {year} {2005})}\BibitemShut {NoStop}%
\bibitem [{\citenamefont {Freyssinet}\ \emph {et~al.}(1983)\citenamefont
	{Freyssinet}, \citenamefont {Torbet}, \citenamefont {Hudry-Clergeon},\ and\
	\citenamefont {Maret}}]{freyssinet1983fibrinogen}%
\BibitemOpen
\bibfield  {author} {\bibinfo {author} {\bibfnamefont {J.}~\bibnamefont
		{Freyssinet}}, \bibinfo {author} {\bibfnamefont {J.}~\bibnamefont {Torbet}},
	\bibinfo {author} {\bibfnamefont {G.}~\bibnamefont {Hudry-Clergeon}}, \ and\
	\bibinfo {author} {\bibfnamefont {G.}~\bibnamefont {Maret}},\ }\href@noop {}
{\bibfield  {journal} {\bibinfo  {journal} {Proceedings of the National
			Academy of Sciences}\ }\textbf {\bibinfo {volume} {80}},\ \bibinfo {pages}
	{1616} (\bibinfo {year} {1983})}\BibitemShut {NoStop}%
\bibitem [{\citenamefont {Torbet}\ and\ \citenamefont
	{Ronziere}(1984)}]{torbet1984magnetic}%
\BibitemOpen
\bibfield  {author} {\bibinfo {author} {\bibfnamefont {J.}~\bibnamefont
		{Torbet}}\ and\ \bibinfo {author} {\bibfnamefont {M.-C.}\ \bibnamefont
		{Ronziere}},\ }\href@noop {} {\bibfield  {journal} {\bibinfo  {journal}
		{Biochem. j}\ }\textbf {\bibinfo {volume} {219}},\ \bibinfo {pages} {1057}
	(\bibinfo {year} {1984})}\BibitemShut {NoStop}%
\bibitem [{\citenamefont {Chionna}\ \emph {et~al.}(2009)\citenamefont
	{Chionna}, \citenamefont {Dwikat}, \citenamefont {Panzarini}, \citenamefont
	{Tenuzzo}, \citenamefont {Carla}, \citenamefont {Verri}, \citenamefont
	{Pagliara}, \citenamefont {Abbro},\ and\ \citenamefont
	{Dini}}]{chionna2009cell}%
\BibitemOpen
\bibfield  {author} {\bibinfo {author} {\bibfnamefont {A.}~\bibnamefont
		{Chionna}}, \bibinfo {author} {\bibfnamefont {M.}~\bibnamefont {Dwikat}},
	\bibinfo {author} {\bibfnamefont {E.}~\bibnamefont {Panzarini}}, \bibinfo
	{author} {\bibfnamefont {B.}~\bibnamefont {Tenuzzo}}, \bibinfo {author}
	{\bibfnamefont {E.}~\bibnamefont {Carla}}, \bibinfo {author} {\bibfnamefont
		{T.}~\bibnamefont {Verri}}, \bibinfo {author} {\bibfnamefont
		{P.}~\bibnamefont {Pagliara}}, \bibinfo {author} {\bibfnamefont
		{L.}~\bibnamefont {Abbro}}, \ and\ \bibinfo {author} {\bibfnamefont
		{L.}~\bibnamefont {Dini}},\ }\href@noop {} {\bibfield  {journal} {\bibinfo
		{journal} {European Journal of Histochemistry}\ }\textbf {\bibinfo {volume}
		{47}},\ \bibinfo {pages} {299} (\bibinfo {year} {2009})}\BibitemShut
{NoStop}%
\bibitem [{\citenamefont {Miyakoshi}(2005)}]{miyakoshi2005effects}%
\BibitemOpen
\bibfield  {author} {\bibinfo {author} {\bibfnamefont {J.}~\bibnamefont
		{Miyakoshi}},\ }\href@noop {} {\bibfield  {journal} {\bibinfo  {journal}
		{Progress in biophysics and molecular biology}\ }\textbf {\bibinfo {volume}
		{87}},\ \bibinfo {pages} {213} (\bibinfo {year} {2005})}\BibitemShut
{NoStop}%
\bibitem [{\citenamefont {Miyakoshi}(2006)}]{miyakoshi2006review}%
\BibitemOpen
\bibfield  {author} {\bibinfo {author} {\bibfnamefont {J.}~\bibnamefont
		{Miyakoshi}},\ }\href@noop {} {\bibfield  {journal} {\bibinfo  {journal}
		{Science and Technology of Advanced Materials}\ }\textbf {\bibinfo {volume}
		{7}},\ \bibinfo {pages} {305} (\bibinfo {year} {2006})}\BibitemShut {NoStop}%
\bibitem [{\citenamefont {Paul}\ \emph {et~al.}(2005)\citenamefont {Paul},
	\citenamefont {Ferl}, \citenamefont {Klingenberg}, \citenamefont {Brooks},
	\citenamefont {Morgan}, \citenamefont {Yowtak},\ and\ \citenamefont
	{Meisel}}]{paul2005strong}%
\BibitemOpen
\bibfield  {author} {\bibinfo {author} {\bibfnamefont {A.-L.}\ \bibnamefont
		{Paul}}, \bibinfo {author} {\bibfnamefont {R.~J.}\ \bibnamefont {Ferl}},
	\bibinfo {author} {\bibfnamefont {B.}~\bibnamefont {Klingenberg}}, \bibinfo
	{author} {\bibfnamefont {J.~S.}\ \bibnamefont {Brooks}}, \bibinfo {author}
	{\bibfnamefont {A.~N.}\ \bibnamefont {Morgan}}, \bibinfo {author}
	{\bibfnamefont {J.}~\bibnamefont {Yowtak}}, \ and\ \bibinfo {author}
	{\bibfnamefont {M.~W.}\ \bibnamefont {Meisel}},\ }in\ \href@noop {} {\emph
	{\bibinfo {booktitle} {Materials Processing in Magnetic Fields}}},\
Vol.~\bibinfo {volume} {1}\ (\bibinfo {organization} {World Scientific},\
\bibinfo {year} {2005})\ pp.\ \bibinfo {pages} {238--242}\BibitemShut
{NoStop}%
\bibitem [{\citenamefont {Teodori}\ \emph {et~al.}(2002)\citenamefont
	{Teodori}, \citenamefont {G{\"o}hde}, \citenamefont {Valente}, \citenamefont
	{Tagliaferri}, \citenamefont {Coletti}, \citenamefont {Perniconi},
	\citenamefont {Bergamaschi}, \citenamefont {Cerella},\ and\ \citenamefont
	{Ghibelli}}]{teodori2002static}%
\BibitemOpen
\bibfield  {author} {\bibinfo {author} {\bibfnamefont {L.}~\bibnamefont
		{Teodori}}, \bibinfo {author} {\bibfnamefont {W.}~\bibnamefont {G{\"o}hde}},
	\bibinfo {author} {\bibfnamefont {M.~G.}\ \bibnamefont {Valente}}, \bibinfo
	{author} {\bibfnamefont {F.}~\bibnamefont {Tagliaferri}}, \bibinfo {author}
	{\bibfnamefont {D.}~\bibnamefont {Coletti}}, \bibinfo {author} {\bibfnamefont
		{B.}~\bibnamefont {Perniconi}}, \bibinfo {author} {\bibfnamefont
		{A.}~\bibnamefont {Bergamaschi}}, \bibinfo {author} {\bibfnamefont
		{C.}~\bibnamefont {Cerella}}, \ and\ \bibinfo {author} {\bibfnamefont
		{L.}~\bibnamefont {Ghibelli}},\ }\href@noop {} {\bibfield  {journal}
	{\bibinfo  {journal} {Cytometry}\ }\textbf {\bibinfo {volume} {49}},\
	\bibinfo {pages} {143} (\bibinfo {year} {2002})}\BibitemShut {NoStop}%
\bibitem [{\citenamefont {Rosen}(2010)}]{rosen2010studies}%
\BibitemOpen
\bibfield  {author} {\bibinfo {author} {\bibfnamefont {A.~D.}\ \bibnamefont
		{Rosen}},\ }\href@noop {} {\bibfield  {journal} {\bibinfo  {journal} {Piers
			Online}\ }\textbf {\bibinfo {volume} {6}},\ \bibinfo {pages} {133} (\bibinfo
	{year} {2010})}\BibitemShut {NoStop}%
\bibitem [{\citenamefont {Higashi}\ \emph {et~al.}(1993)\citenamefont
	{Higashi}, \citenamefont {Yamagishi}, \citenamefont {Takeuchi}, \citenamefont
	{Kawaguchi}, \citenamefont {Sagawa}, \citenamefont {Onishi},\ and\
	\citenamefont {Date}}]{higashi1993orientation}%
\BibitemOpen
\bibfield  {author} {\bibinfo {author} {\bibfnamefont {T.}~\bibnamefont
		{Higashi}}, \bibinfo {author} {\bibfnamefont {A.}~\bibnamefont {Yamagishi}},
	\bibinfo {author} {\bibfnamefont {T.}~\bibnamefont {Takeuchi}}, \bibinfo
	{author} {\bibfnamefont {N.}~\bibnamefont {Kawaguchi}}, \bibinfo {author}
	{\bibfnamefont {S.}~\bibnamefont {Sagawa}}, \bibinfo {author} {\bibfnamefont
		{S.}~\bibnamefont {Onishi}}, \ and\ \bibinfo {author} {\bibfnamefont
		{M.}~\bibnamefont {Date}},\ }\href@noop {} {\bibfield  {journal} {\bibinfo
		{journal} {Blood}\ }\textbf {\bibinfo {volume} {82}},\ \bibinfo {pages}
	{1328} (\bibinfo {year} {1993})}\BibitemShut {NoStop}%
\bibitem [{\citenamefont {Hong}(1995)}]{hong1995magnetic}%
\BibitemOpen
\bibfield  {author} {\bibinfo {author} {\bibfnamefont {F.~T.}\ \bibnamefont
		{Hong}},\ }\href@noop {} {\bibfield  {journal} {\bibinfo  {journal}
		{Biosystems}\ }\textbf {\bibinfo {volume} {36}},\ \bibinfo {pages} {187}
	(\bibinfo {year} {1995})}\BibitemShut {NoStop}%
\bibitem [{\citenamefont {Tenuzzo}\ \emph {et~al.}(2009)\citenamefont
	{Tenuzzo}, \citenamefont {Vergallo},\ and\ \citenamefont
	{Dini}}]{tenuzzo2009effect}%
\BibitemOpen
\bibfield  {author} {\bibinfo {author} {\bibfnamefont {B.}~\bibnamefont
		{Tenuzzo}}, \bibinfo {author} {\bibfnamefont {C.}~\bibnamefont {Vergallo}}, \
	and\ \bibinfo {author} {\bibfnamefont {L.}~\bibnamefont {Dini}},\ }\href@noop
{} {\bibfield  {journal} {\bibinfo  {journal} {Tissue and Cell}\ }\textbf
	{\bibinfo {volume} {41}},\ \bibinfo {pages} {169} (\bibinfo {year}
	{2009})}\BibitemShut {NoStop}%
\bibitem [{\citenamefont {N{\'e}el}(1948)}]{neel1948proprietes}%
\BibitemOpen
\bibfield  {author} {\bibinfo {author} {\bibfnamefont {L.}~\bibnamefont
		{N{\'e}el}},\ }in\ \href@noop {} {\emph {\bibinfo {booktitle} {Annales de
			Physique}}},\ Vol.~\bibinfo {volume} {3}\ (\bibinfo {organization} {EDP
	SCIENCES SA 17, AVE DU HOGGAR, PA COURTABOEUF, BP 112, F-91944 LES ULIS CEDEX
	A, FRANCE},\ \bibinfo {year} {1948})\ pp.\ \bibinfo {pages}
{137--198}\BibitemShut {NoStop}%
\bibitem [{\citenamefont {Peskin}\ \emph {et~al.}(1993)\citenamefont {Peskin},
	\citenamefont {Odell},\ and\ \citenamefont {Oster}}]{peskin1993cellular}%
\BibitemOpen
\bibfield  {author} {\bibinfo {author} {\bibfnamefont {C.~S.}\ \bibnamefont
		{Peskin}}, \bibinfo {author} {\bibfnamefont {G.~M.}\ \bibnamefont {Odell}}, \
	and\ \bibinfo {author} {\bibfnamefont {G.~F.}\ \bibnamefont {Oster}},\
}\href@noop {} {\bibfield  {journal} {\bibinfo  {journal} {Biophysical
		journal}\ }\textbf {\bibinfo {volume} {65}},\ \bibinfo {pages} {316}
(\bibinfo {year} {1993})}\BibitemShut {NoStop}%
\bibitem [{\citenamefont {Wang}\ and\ \citenamefont
	{Oster}(2002)}]{wang2002ratchets}%
\BibitemOpen
\bibfield  {author} {\bibinfo {author} {\bibfnamefont {H.}~\bibnamefont
		{Wang}}\ and\ \bibinfo {author} {\bibfnamefont {G.}~\bibnamefont {Oster}},\
}\href@noop {} {\bibfield  {journal} {\bibinfo  {journal} {Applied Physics
		A}\ }\textbf {\bibinfo {volume} {75}},\ \bibinfo {pages} {315} (\bibinfo
{year} {2002})}\BibitemShut {NoStop}%
\bibitem [{\citenamefont {Tomkiewicz}\ \emph {et~al.}(2007)\citenamefont
	{Tomkiewicz}, \citenamefont {Nouwen},\ and\ \citenamefont
	{Driessen}}]{tomkiewicz2007pushing}%
\BibitemOpen
\bibfield  {author} {\bibinfo {author} {\bibfnamefont {D.}~\bibnamefont
		{Tomkiewicz}}, \bibinfo {author} {\bibfnamefont {N.}~\bibnamefont {Nouwen}},
	\ and\ \bibinfo {author} {\bibfnamefont {A.~J.}\ \bibnamefont {Driessen}},\
}\href@noop {} {\bibfield  {journal} {\bibinfo  {journal} {FEBS letters}\
}\textbf {\bibinfo {volume} {581}},\ \bibinfo {pages} {2820} (\bibinfo {year}
{2007})}\BibitemShut {NoStop}%
\bibitem [{\citenamefont {Shaw}\ \emph {et~al.}(2014)\citenamefont {Shaw},
	\citenamefont {Raja},\ and\ \citenamefont {Dasgupta}}]{shaw2014modulation}%
\BibitemOpen
\bibfield  {author} {\bibinfo {author} {\bibfnamefont {J.}~\bibnamefont
		{Shaw}}, \bibinfo {author} {\bibfnamefont {S.~O.}\ \bibnamefont {Raja}}, \
	and\ \bibinfo {author} {\bibfnamefont {A.~K.}\ \bibnamefont {Dasgupta}},\
}\href@noop {} {\bibfield  {journal} {\bibinfo  {journal} {Cancer
		Nanotechnology}\ }\textbf {\bibinfo {volume} {5}},\ \bibinfo {pages} {1}
(\bibinfo {year} {2014})}\BibitemShut {NoStop}%
\bibitem [{\citenamefont {Changeux}\ \emph {et~al.}(1967)\citenamefont
	{Changeux}, \citenamefont {Thi{\'e}ry}, \citenamefont {Tung},\ and\
	\citenamefont {Kittel}}]{changeux1967cooperativity}%
\BibitemOpen
\bibfield  {author} {\bibinfo {author} {\bibfnamefont {J.-P.}\ \bibnamefont
		{Changeux}}, \bibinfo {author} {\bibfnamefont {J.}~\bibnamefont
		{Thi{\'e}ry}}, \bibinfo {author} {\bibfnamefont {Y.}~\bibnamefont {Tung}}, \
	and\ \bibinfo {author} {\bibfnamefont {C.}~\bibnamefont {Kittel}},\
}\href@noop {} {\bibfield  {journal} {\bibinfo  {journal} {Proceedings of the
		National Academy of Sciences of the United States of America}\ }\textbf
{\bibinfo {volume} {57}},\ \bibinfo {pages} {335} (\bibinfo {year}
{1967})}\BibitemShut {NoStop}%
\bibitem [{\citenamefont {Yue}\ \emph {et~al.}(2012)\citenamefont {Yue},
	\citenamefont {Lee}, \citenamefont {Afkhami},\ and\ \citenamefont
	{Renardy}}]{yue2012motion}%
\BibitemOpen
\bibfield  {author} {\bibinfo {author} {\bibfnamefont {P.}~\bibnamefont
		{Yue}}, \bibinfo {author} {\bibfnamefont {S.}~\bibnamefont {Lee}}, \bibinfo
	{author} {\bibfnamefont {S.}~\bibnamefont {Afkhami}}, \ and\ \bibinfo
	{author} {\bibfnamefont {Y.}~\bibnamefont {Renardy}},\ }\href@noop {}
{\bibfield  {journal} {\bibinfo  {journal} {Acta Mechanica}\ }\textbf
	{\bibinfo {volume} {223}},\ \bibinfo {pages} {505} (\bibinfo {year}
	{2012})}\BibitemShut {NoStop}%
\bibitem [{\citenamefont {Chen}\ \emph {et~al.}(2013)\citenamefont {Chen},
	\citenamefont {Byvank}, \citenamefont {Vieira},\ and\ \citenamefont
	{Sooryakumar}}]{chen2013magnetic}%
\BibitemOpen
\bibfield  {author} {\bibinfo {author} {\bibfnamefont {A.}~\bibnamefont
		{Chen}}, \bibinfo {author} {\bibfnamefont {T.}~\bibnamefont {Byvank}},
	\bibinfo {author} {\bibfnamefont {G.~B.}\ \bibnamefont {Vieira}}, \ and\
	\bibinfo {author} {\bibfnamefont {R.}~\bibnamefont {Sooryakumar}},\
}\href@noop {} {\bibfield  {journal} {\bibinfo  {journal} {Magnetics, IEEE
		Transactions on}\ }\textbf {\bibinfo {volume} {49}},\ \bibinfo {pages} {300}
(\bibinfo {year} {2013})}\BibitemShut {NoStop}%
\bibitem [{\citenamefont {Lambert}\ \emph {et~al.}(2013)\citenamefont
	{Lambert}, \citenamefont {Chen}, \citenamefont {Cheng}, \citenamefont {Li},
	\citenamefont {Chen},\ and\ \citenamefont {Nori}}]{lambert2013quantum}%
\BibitemOpen
\bibfield  {author} {\bibinfo {author} {\bibfnamefont {N.}~\bibnamefont
		{Lambert}}, \bibinfo {author} {\bibfnamefont {Y.-N.}\ \bibnamefont {Chen}},
	\bibinfo {author} {\bibfnamefont {Y.-C.}\ \bibnamefont {Cheng}}, \bibinfo
	{author} {\bibfnamefont {C.-M.}\ \bibnamefont {Li}}, \bibinfo {author}
	{\bibfnamefont {G.-Y.}\ \bibnamefont {Chen}}, \ and\ \bibinfo {author}
	{\bibfnamefont {F.}~\bibnamefont {Nori}},\ }\href@noop {} {\bibfield
	{journal} {\bibinfo  {journal} {Nature Physics}\ }\textbf {\bibinfo {volume}
		{9}},\ \bibinfo {pages} {10} (\bibinfo {year} {2013})}\BibitemShut {NoStop}%
\bibitem [{\citenamefont {Bhattacharya}\ \emph {et~al.}(2014)\citenamefont
	{Bhattacharya}, \citenamefont {Chakraborty}, \citenamefont {Raja},
	\citenamefont {Ghosh}, \citenamefont {Dasgupta},\ and\ \citenamefont
	{Dasgupta}}]{bhattacharya2014static}%
\BibitemOpen
\bibfield  {author} {\bibinfo {author} {\bibfnamefont {A.}~\bibnamefont
		{Bhattacharya}}, \bibinfo {author} {\bibfnamefont {M.}~\bibnamefont
		{Chakraborty}}, \bibinfo {author} {\bibfnamefont {S.~O.}\ \bibnamefont
		{Raja}}, \bibinfo {author} {\bibfnamefont {A.}~\bibnamefont {Ghosh}},
	\bibinfo {author} {\bibfnamefont {M.}~\bibnamefont {Dasgupta}}, \ and\
	\bibinfo {author} {\bibfnamefont {A.~K.}\ \bibnamefont {Dasgupta}},\
}\href@noop {} {\bibfield  {journal} {\bibinfo  {journal} {Photochemical \&
		Photobiological Sciences}\ } (\bibinfo {year} {2014})}\BibitemShut {NoStop}%
\bibitem [{\citenamefont {Panitchayangkoon}\ \emph {et~al.}(2010)\citenamefont
	{Panitchayangkoon}, \citenamefont {Hayes}, \citenamefont {Fransted},
	\citenamefont {Caram}, \citenamefont {Harel}, \citenamefont {Wen},
	\citenamefont {Blankenship},\ and\ \citenamefont
	{Engel}}]{panitchayangkoon2010long}%
\BibitemOpen
\bibfield  {author} {\bibinfo {author} {\bibfnamefont {G.}~\bibnamefont
		{Panitchayangkoon}}, \bibinfo {author} {\bibfnamefont {D.}~\bibnamefont
		{Hayes}}, \bibinfo {author} {\bibfnamefont {K.~A.}\ \bibnamefont {Fransted}},
	\bibinfo {author} {\bibfnamefont {J.~R.}\ \bibnamefont {Caram}}, \bibinfo
	{author} {\bibfnamefont {E.}~\bibnamefont {Harel}}, \bibinfo {author}
	{\bibfnamefont {J.}~\bibnamefont {Wen}}, \bibinfo {author} {\bibfnamefont
		{R.~E.}\ \bibnamefont {Blankenship}}, \ and\ \bibinfo {author} {\bibfnamefont
		{G.~S.}\ \bibnamefont {Engel}},\ }\href@noop {} {\bibfield  {journal}
	{\bibinfo  {journal} {Proceedings of the National Academy of Sciences}\
	}\textbf {\bibinfo {volume} {107}},\ \bibinfo {pages} {12766} (\bibinfo
	{year} {2010})}\BibitemShut {NoStop}%
\bibitem [{\citenamefont {Jordan}\ \emph {et~al.}(1996)\citenamefont {Jordan},
	\citenamefont {Schallhorn},\ and\ \citenamefont
	{Wurm}}]{jordan1996transfecting}%
\BibitemOpen
\bibfield  {author} {\bibinfo {author} {\bibfnamefont {M.}~\bibnamefont
		{Jordan}}, \bibinfo {author} {\bibfnamefont {A.}~\bibnamefont {Schallhorn}},
	\ and\ \bibinfo {author} {\bibfnamefont {F.~M.}\ \bibnamefont {Wurm}},\
}\href@noop {} {\bibfield  {journal} {\bibinfo  {journal} {Nucleic acids
		research}\ }\textbf {\bibinfo {volume} {24}},\ \bibinfo {pages} {596}
(\bibinfo {year} {1996})}\BibitemShut {NoStop}%
\end{thebibliography}
\end{document}